# High-repetition-rate terahertz and ultraviolet radiation for high-throughput ultrafast electron diffraction

A. Ryabov[1], K. Amini[1,*]

[1]Max-Born-Institut, Max-Born-Str. 2A, 12489, Berlin, Germany.
*Corresponding author email: kasra.amini@mbi-berlin.de

**Abstract**

Scaling femtosecond terahertz (THz) and ultraviolet (UV) sources to high repetition rates is essential for high-throughput ultrafast spectroscopy and imaging applications. Yet, their efficient generation at high average power remains limited by thermal effects, phase-matching constraints, and material damage. Here, we demonstrate broadband THz and UV generation driven by a common Yb:KGW laser operating from at 40-600 kHz. THz radiation is produced by optical rectification in stoichiometric $MgO:LiNbO_3$ using a line-focus geometry, yielding single-cycle pulses of 55-92 nJ energy with peak electric fields of 37-90 kV/cm. Electro-optic sampling and beam-quality measurements reveal tunable control between central frequency, bandwidth and field amplitude by translating the generation region transversely within the crystal. Using shorter pump pulses preserves THz conversion efficiency, while longer pulses at 100 kHz reduce THz output by up to a factor of four due to cumulative thermal effects. Femtosecond 257.5 nm UV pulses are generated by cascaded fourth-harmonic generation in $\beta$-barium borate with conversion efficiencies exceeding 10% at 40 kHz and stable operation up to 600 kHz. These results demonstrate a compact, thermally robust platform for high-average-power nonlinear conversion and are directly relevant to next-generation high-repetition-rate ultrafast electron diffraction and spectroscopy systems.

**Introduction**

High-repetition-rate generation of terahertz (THz) and ultraviolet (UV) radiation has become increasingly important across ultrafast optics, imaging, and spectroscopy. The ability to deliver femtosecond, high-average-power beams in these spectral regions underpins a wide variety of experiments in atomic and molecular photophysics[1,2], condensed matter physics[3–5], and materials science[6]. Beyond spectroscopy, THz and UV pulses are key enablers for compact ultrafast electron sources[7]: low-energy (pJ-nJ) UV pulses drive photoemission for electron probes[8], nJ-level THz fields enable temporal compression and streaking of relativistic electron bunches[9], and µJ-level UV pulses are routinely used for optical excitation of target samples. Despite their central role in ultrafast science, efficient scaling of both THz and UV generation beyond tens of kilohertz (kHz) remains a major experimental challenge.

Ytterbium-based femtosecond laser systems now routinely deliver µJ-mJ pulses at tens-to-hundreds of kHz repetition rates with average powers exceeding 100 W,[10–12] and are beginning to move towards the kilowatt[13,14,12] and higher[15,16] level. These Ytterbium (Yb) sources have opened the possibility to high-throughput nonlinear conversion at unprecedented average powers, provided that thermal absorption, phase-matching, and damage thresholds are carefully managed. Yet, generating high-energy THz and UV pulses at 100 kHz or higher is still difficult, in part because limited work exists on nonlinear conversion under high thermal and average-power loads.

THz generation[17–30] at the µJ level[18,31] has been demonstrated through non-collinear optical rectification using tilted-pulse-front schemes,[18,19,25] which achieve high conversion efficiency but involve complex geometries and face significant thermal challenges at high average power[32]. In contrast, collinear geometries employing line focusing on Mg-doped $LiNbO_3$ offer a simpler and more robust route to THz generation by optical rectification[20]. This approach reduces the thermal load while producing broadband single-cycle THz radiation at tens of kHz[20]. For UV generation, several strategies have been pursued, including nonlinear upconversion in birefringent crystals[33–35] and high-pressure laser-fabricated glass cells[36,37], resonant dispersive-wave emission in gas-filled fibers[38–41], and four-wave mixing in noble gases[42,43]. With Yb-based lasers, whose fundamental output is typically centered at 1030 nm, direct third harmonic generation does not reach the UV regime as in Ti:sapphire-based systems. Cascaded fourth-harmonic generation (FHG)[44] of 1030 nm in birefringent β-barium borate (BBO) crystals provides a compact and efficient route to 257.5 nm UV generation,[45] but its scaling beyond 100 kHz with sub-picosecond pump durations is largely unexplored.

Here, in this work, we present a comprehensive experimental study of high-repetition-rate THz and UV generation driven by a common Yb:KGW thin disk laser system (Light Conversion CARBIDE, 1030 nm, 320 fs FWHM, 2 mJ, 80 W, 40 kHz tunable up to 2 MHz). THz radiation is produced by optical rectification in stoichiometric MgO:LiNbO$_3$ using a line-focus geometry, while 257.5 nm UV pulses are generated through cascaded FHG in BBO. We systematically examine scaling behavior of both processes with repetition rate, pulse duration, and pump power – spanning from 40 kHz to up to 100 kHz for THz generation and up to 600 kHz for UV generation. We observe efficient, stable THz emission up to 92 nJ with single-cycle waveforms, and UV generation from the nJ to pJ regime suitable for generating electron probes, establishing a scalable platform for high-average-power ultrafast sources, for example ultrafast electron diffraction (UED)[2,7,46–48,8]. We first analyze THz generation at different repetition rates as well as pump pulse duration and powers, followed by repetition-scaling of cascaded FHG conversion in BBOs.

Figure 1 shows the optical schematic of the THz generation setup based on Ref. [20]. A cylindrical telescope and plano-convex lens produce a line focus of 1030 nm radiation by magnifying in the $y$-axis and focussing in the $x$-axis (3.4 mm × 56 μm 1/e$^2$ radius). The $y$-polarized line focus impinges on a thin MgO-doped stoichiometric LiNbO$_3$ slab (1 × 10 × 10 mm$^3$). Both the line focus geometry and stoichiometric doping reduce the thermal load on the crystal. The 1030 nm pulse energy was varied up to 225 μJ at both 40 kHz and 100 kHz (*i.e.*, 9 W and 22 W average power, respectively), and the pulse duration was adjusted between 320 fs and 1 ps using a grating compressor. A silicon prism (40.4° apex) is optically contacted to the output face of the LiNbO$_3$ slab to efficiently outcouple the generated THz emission[20]. The refractive-index matching between LiNbO$_3$ ($n = 5.0$ at 0.5 THz) and silicon ($n = 3.4$) minimizes total internal reflection losses at the interface and allows near-Cherenkov phase-matched extraction of THz frequencies up to ~2 THz. The optical contact between the prism and the crystal eliminates any air gap, thereby suppressing Fresnel reflection losses and avoiding Fabry-Pérot interference distortion of the temporal THz waveform. The 1030 nm pump pulse before the cylindrical lenses was mechanically chopped, and the average power of the outcoupled THz beam, generated in a collinear geometry, was measured using a pyroelectric detector (Gentec-EO), while the spatial profile was recorded with a THz beam profiler (Pyrocam-III, Spiricon Inc). Electro-optic sampling (EOS) measurements were performed in an EOS crystal (ZnTe, 0.5 mm), using the THz radiation as a pump and a weak 1030 nm pulse as a probe to retrieve the temporal electric-field waveform. The transmitted 1030 nm probe beam passes through a quarter waveplate, a Wollaston prism and onto a balanced photodiode detector for EOS detection. These EOS measurements enabled us to determine the THz pulse contrast (*i.e.*, the ratio between the $+x$ Cherenkov pulse to the $-x$ pulse) as well as the delay between the two Cherenkov pulses.

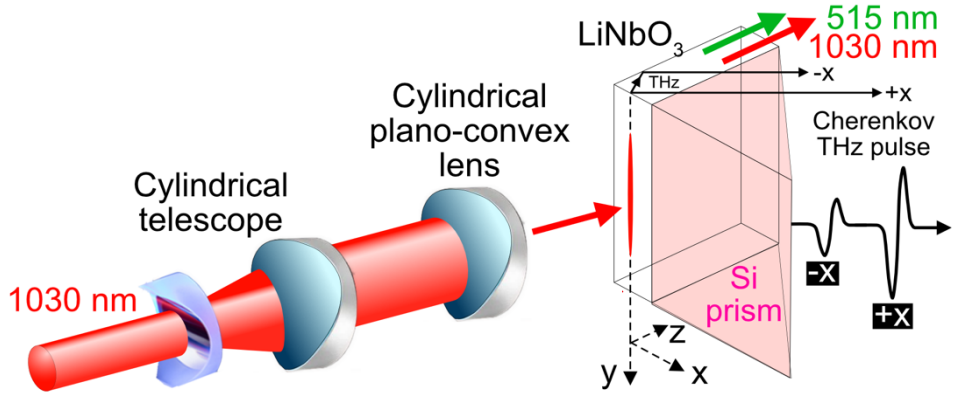

**FIG. 1.** Experimental setup of THz generation. A cylindrical telescope and plano-convex lens form a $y$-polarized line focus on a LiNbO$_3$ slab, generating THz radiation that is efficiently out-coupled through an optically contacted silicon prism. The $-x$ and $+x$ Cherenkov THz pulses are shown.

Figure 2 summarizes the emitted THz pulse energy as a function of input 1030 nm energy for different pulse durations (320 fs and 1000 fs) and repetition rates (40 kHz and 100 kHz) at various $x$-positions of the LiNbO$_3$ slab. At the slab center ($x = 0.0$ mm), up to 55 nJ of $+x$ Cherenkov THz radiation is generated using 214 µJ of total input 1030 nm energy (see Fig. 2a), corresponding to a conversion efficiency of $2.6 \times 10^{-4}$, close to the optimal efficiency reported by Tsarev et al.[20]. Translating the line focus along the $x$-axis away from the center leads to a gradual increase in emitted THz energy (see Fig. 2c). As the generation region moves towards the exit face ($+x$ Cherenkov-emission side, see Fig. 1), the internal propagation distance of the THz field decreases, reducing both absorption and total internal reflection losses. Consequently, a larger portion of the THz spectrum is transmitted through the crystal and collected within the fixed numerical aperture (NA) of the THz optics, including higher-frequency components that are otherwise absorbed deeper inside the LiNbO$_3$ slab.

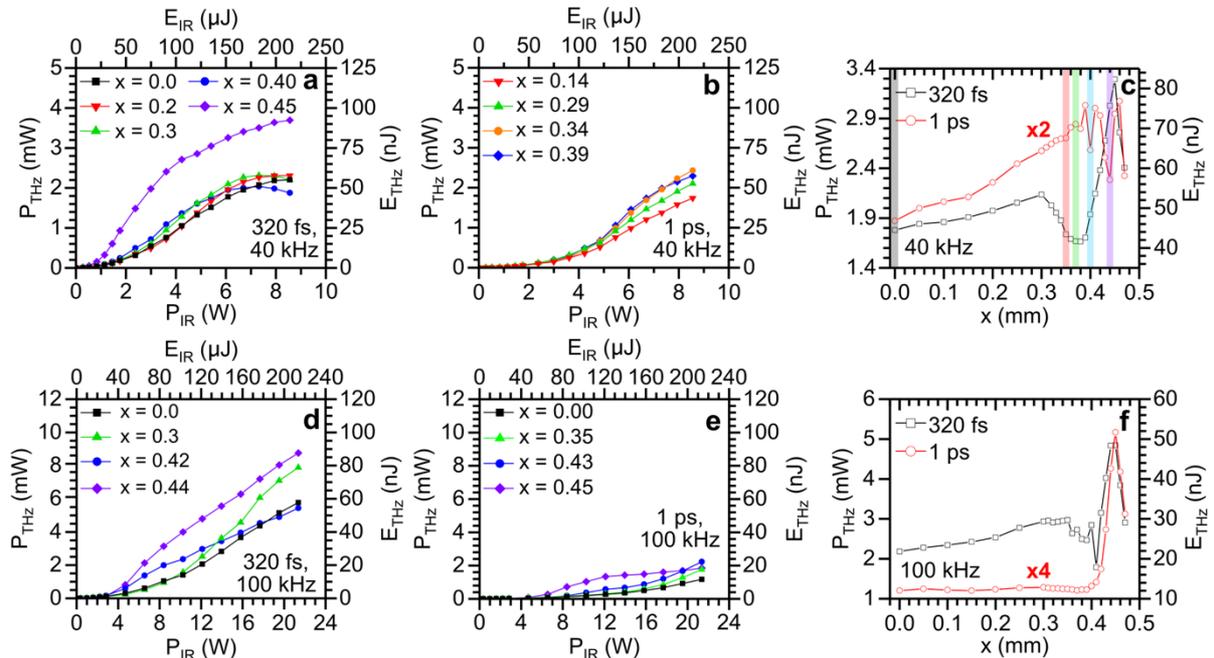

**FIG. 2.** Emitted THz radiation at 40 kHz (top) and 100 kHz (bottom). (a-b) Measured THz output as a function of 1030 nm infrared (IR) input with a duration of 320 fs (a) and 1 ps (b). (c) Measured THz output as a function of the $x$-position of the 1030 nm line focus at the LiNbO$_3$ slab at 40 kHz for both 320 fs (black squares) and 1 ps (red circles) input pulse duration. (d-f) Same as (a-c) but for 100 kHz. The measured THz radiation is reported only for the $+x$ Cherenkov pulse and chopper-open intervals.

A local minimum near $x \approx 0.38$ mm appears because the shorter 320 fs pump pulse (see green shaded and black distribution in Fig. 2c) produces a broader THz radiation, with higher frequencies emitted at larger internal Cherenkov angles. Part of this radiation exits the crystal at angles exceeding the collection NA, leading to a reduced detected signal. When the generation region is positioned within approximately one THz attenuation length of the exit face ($x > 0.40$ mm), absorption and reflection losses are strongly supressed, allowing the THz field to couple directly into free space. This enhanced out-coupling more than compensates for NA collection losses, resulting in a sharp increase in detected THz energy, allowing a greater fraction of both low- and high-frequency components to escape the crystal. At the crystal edge ($x = 0.45$ mm), the emitted THz energy reaches 92 nJ, which is 67% higher than at the slab center, corresponding to a conversion efficiency of $4.3 \times 10^{-4}$.

At 100 kHz with a 320-fs pulse (see Fig. 2d), the position-dependent behavior persists, though the reduction at around $x = 0.38$ mm is less pronounced. This suggests that there is a modest thermally induced softening of the THz spectrum and a subsequent reduction of the internal Cherenkov angle; the resulting improvement in collection efficiency partially offsets the angular overfilling losses that dominate at 40 kHz. Even at 22 W average pump power, efficient THz generation is maintained with pulse energies of 54-88 nJ at 100 kHz, demonstrating that optical rectification in $LiNbO_3$ remains robust and scalable to high repetition rates.

To further examine the frequency-dependence behavior, THz generation was investigated using a longer 1 ps pump pulse (see Figs. 2b and 2e), comparable to the conditions of Tsarev et al.[20]. At 40 kHz, the longer pulse of 1 ps produces a narrower, low-frequency THz spectrum with longer coherence lengths and smaller internal emission angles. As the generation region moves away from the slab center, the shorter internal path length reduces absorption losses, maintaining efficient out-coupling despite the lower peak intensity. Near the exit face ($x > 0.4$ mm), however, the reduced interaction volume truncates the coherent buildup of the THz field, introducing destructive interference at the boundary and resulting in a sharp decrease in detected energy – an effect observed consistently across all datasets (Figs. 2c,f). In general, at 40 kHz, THz pulse energies of 44-61 nJ are generated with a 1 ps pulse.

In contrast, at 100 kHz with a 1 ps pump pulse, the THz output (12-22 nJ) decreases by a factor of four compared to all other measured configurations (44-92 nJ). This decrease cannot be solely attributed to the longer pulse duration, since at 40 kHz the THz yield remains essentially unchanged when increasing the pulse duration from 320 fs to 1 ps. The lower THz conversion efficiency at 100 kHz instead originates from cumulative thermal effects within the $LiNbO_3$ slab (e.g., increased THz absorption, thermal lensing, thermo-optic dephasing) that distort the phase-matching and limit the effective nonlinear interaction length. Nonetheless, using the shorter 320 fs pulse at 100 kHz still produces efficient THz generation, suggesting that a higher peak intensity because of the shorter pulse and a broader pump bandwidth drive stronger instantaneous nonlinear polarization (*i.e.*, optical rectification) before thermal gradients accumulate, emphasizing the importance of short-pulse excitation at high average power.

EOS measurements at selected $x$-positions of the $LiNbO_3$ crystal (see shaded regions in Fig. 2c) provide direct confirmation of the position-dependent spectral and temporal behavior inferred from Fig. 2. At the slab center, the THz field is single-cycle with a spectral peak at 0.52 THz (see black solid distribution in Figs. 3a-b). Moving away from the slab center introduces additional oscillations and higher-frequency components (see grey diagonal line in Fig. 3b), leading to a broader frequency distribution with the appearance of spectral content beyond 1 THz (vertical dashed line in Fig. 3b). Near the exit face ($x = 0.45$ mm), the mean frequency increases to 0.67 THz. Even though the THz conversion efficiency reaches its maximum at the exit face, the amplitude of the first half-cycle decreases by 23% compared to that generated at the slab center. This reduction is undesirable for THz-based electron compression and streaking experiments, which rely on the steepest single-cycle field to achieve the highest THz field strength, and therefore, strongest electron deflection. EOS measurements confirm that the longer pulse of 1 ps at 40 kHz leads to a lower-frequency, single-cycle THz waveform with a central frequency of 0.36 THz (see black dotted distributions in Fig. 3a-b) that is close to what was reported by Tsarev et al. (0.26 THz) under similar pump conditions.

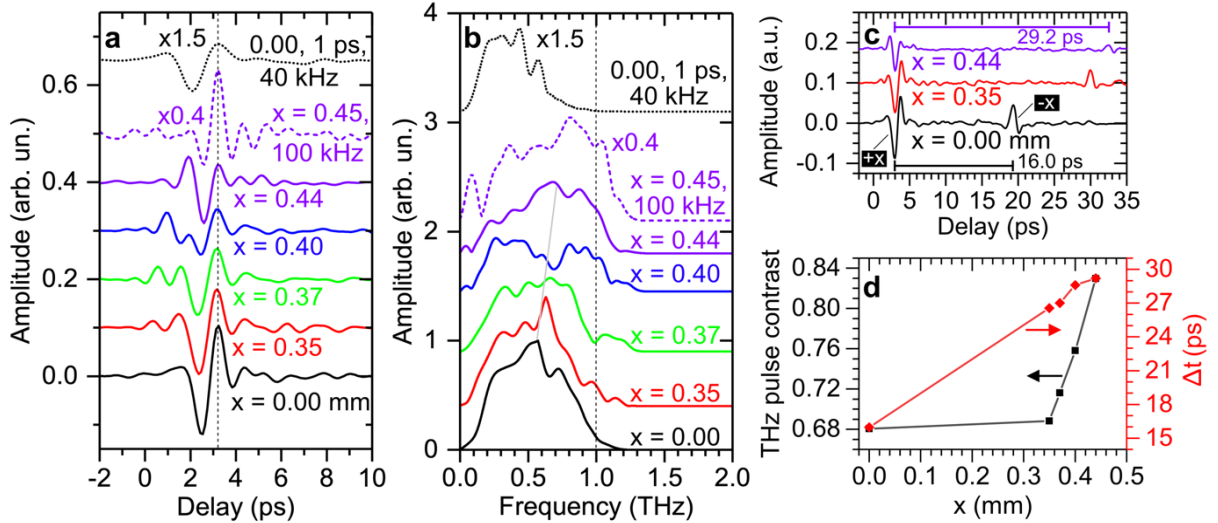

**FIG. 3.** Electro-optic sampling (EOS) measurements of emitted THz radiation. (a) EOS measurements at different $x$-positions of the line focus on LiNbO$_3$ slab at 40 kHz (solid lines) and 100 kHz (dashed line). A scan at 40 kHz with 1 ps duration (black dotted line) is also shown. (b) Frequency spectrum extracted from the EOS measurements shown in panel (a). The shift in central frequency at 40 kHz is indicated by the grey diagonal line. The spectral content at and above 1 THz is indicated by the black dashed vertical line. (c) EOS measurements at 40 kHz for three selected $x$-positions on LiNbO$_3$ slab showing the $+x$ and $-x$ Cherenkov pulses. The time delay between the two Cherenkov pulses is indicated by horizontal bar lines. (d) THz pulse contrast of the $+x$ to the $-x$ Cherenkov pulse (black) and the delay between two pulses (red) as a function of the $x$-positions of the line focus on LiNbO$_3$ slab.

EOS measurements at 100 kHz exhibit similar features near the exit face, such as the presence of higher-frequency components in both the temporal and spectral domains (see purple dashed in Figs. 3a-b) while maintaining a comparable single-cycle waveform. These observations further confirm that with a 320-fs pulse, increasing repetition rate does not degrade the waveform or its spectral content. However, they also highlight that conversion efficiency alone is not a sufficient performance metric – accurate knowledge of the temporal and spectral field structure is essential. For THz-based electron measurements, generation near the crystal center therefore remains optimal. In contrast, for THz spectroscopy studies, positioning the generation region near the exit face can be advantageous due to the enhanced high-frequency spectral content.

It is also important to note that the pump pulse parameters also influence the optimal line-focus geometry, the dimensions of the LiNbO$_3$ slab, and the effective interaction length within the LiNbO$_3$ slab. For our shorter and lower-energy pulses (320 fs, 220 µJ) compared to Tsarev et al.[20] (1 ps, 300 µJ), their model predicts optimum focusing parameters of $w_x = 18$ µm and $w_y = 22.7$ mm with an ideal LiNbO$_3$ crystal $z$-length of 2.2 mm, corresponding to a peak frequency of 0.83 THz. Considering our 10 mm $z$-length of our slab, the Rayleigh-length constraint necessitates a looser focus ($w_x = 39 - 46$ µm) to maintain uniform pump intensity across the interaction region. This broader focus increases the effective THz single-cycle pulse duration to 645 fs and shifts the predicted spectral peak to ~0.50 THz, in excellent agreement with the EOS-measured value of 0.52 THz at the crystal center ($x = 0.0$ mm).

In the time-domain, the EOS data additionally reveal the presence of a secondary THz waveform corresponding to the $-x$ Cherenkov pulse (see Fig. 3c) with opposite phase, as previously briefly reported in Tsarev *et al*. As the generation region is translated from the slab center (x = 0.0 mm) to the exit face (x = 0.44 mm), this backward-emitted THz component experiences strong absorption, leading to a noticeable reduction in its amplitude. Meanwhile, the temporal delay between the $+x$ and $-x$ Cherenkov pulses increase from 16.0 ps to 29.2 ps (red distribution in Fig. 3d). The corresponding THz pulse contrast, defined as the ratio of the $+x$ to $-x$ Cherenkov pulse amplitudes, remains approximately constant (~0.68) up to $x = 0.35$ mm, above which it increases sharply (~0.83), confirming the progressive attenuation of the reflected $-x$ Cherenkov pulse when the generation is close to the $+x$ Cherenkov exit face.

Beam profiling of the THz radiation at 100 kHz (Fig. 4) shows a 1.15 mm FWHM diameter at the slab center (see inset of Fig. 4a), reducing to 0.67 mm near the exit face (see inset of Fig. 4b), consistent with increased higher-frequency components producing tighter focusing. From their measured THz pulse energies and beam waists, the peak electric field strength in air[20] is estimated to reach up to 37 kV/cm at the center and 90 kV/cm near the exit face. These values exceed the 30 kV/cm reported by Tsarev *et al.* at 0.265 THz source, primarily because our higher THz frequency permits tighter focusing for comparable pulse energies, and our source delivers single-cycle pulses with higher energy without loss of temporal fidelity.

Beam-quality M[2] measurements corroborate these findings. THz radiation at the slab center (see Fig. 4) gives $M_x^2 = 1.34$ and $M_y^2 = 1.73$ (Fig. 4a), improving to $M_x^2 = 1.02$ and $M_y^2 = 1.15$ near the exit face (Fig. 4b). The lower M[2] values indicate a more Gaussian-like THz beam and reduced astigmatism, consistent with the more symmetric spatial profiles shown in Fig. 4b with nearly diffraction-limited beam quality. This improvement is due to the reduced propagation distance and fewer internal reflections at the exit face, which suppress wavefront distortion and angular dispersion accumulated within the crystal.

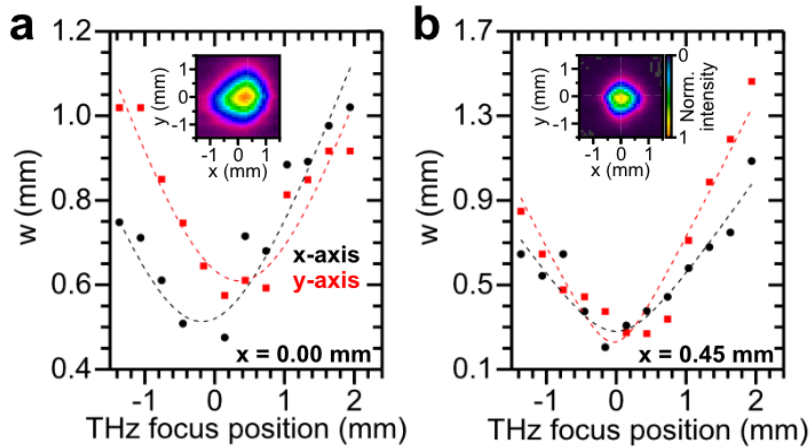

**FIG. 4.** (a-b) Beam profile and M[2] measurements of THz radiation at the center of the LiNbO$_3$ slab (a) and close to the exit face (b) along the $x$- (black) and $y$-axes (red). Measured (squares, circles) and fitted (dashed lines) 1/e$^2$ radii are shown with inset images of the beam profile at the THz focus position.

Overall, the EOS and beam-profiling measurements reveal that the temporal, spectral, and spatial properties of the emitted THz field depend sensitively on the generation position within the LiNbO$_3$ slab. Generation near the exit face enhances bandwidth and conversion efficiency through improved out-coupling, whereas generation at the slab center yields single-cycle fields with higher peak amplitudes – conditions favorable for THz-driven electron compression and streaking. This trade-off between field strength and bandwidth highlights the tunability of the LiNbO$_3$ source and its scalability to high average powers.

Building on these results, we next investigate the generation of UV radiation through cascaded fourth-harmonic generation (FHG) using the same Yb-based laser platform, extending the study of high-repetition-rate nonlinear conversion toward photon energies relevant for photoemission of electron pulses and sample excitation, particularly relevant for high-repetition-rate UED.

*Fourth-harmonic generation of UV radiation*

Figure 5a shows the cascaded two-stage SHG scheme corresponding to FHG at 1030 nm. The 1030 nm fundamental pulse (<60 mW, <1.5 µJ, 320 fs) passes through a 1:1 Keplerian telescope with a BBO crystal positioned at the focus (<30 µm $1/e^2$ diameter) for type-I SHG ($\theta = 23.4°$), generating 515 nm radiation. A harmonic separator reflects the residual 1030 nm while transmitting the 515 nm beam into a second 1:1 Keplerian telescope, where a second BBO produces type-I SHG of 257.5 nm radiation. The UV output is separated using UV harmonic separators and directed by UV high reflector mirrors to diagnostics. The measured spectra and spatial profiles of the 1030 nm, 515 nm and 257.5 nm beams are shown in Figs. 5b-d. The spectra yield FWHM bandwidths of 5.6 nm, 1.75 nm, and 0.68 nm at 1030 nm, 515 nm, and 257.5 nm, respectively, corresponding to Fourier-transform-limited (FTL) durations of 280 fs, 225 fs, and 144 fs, respectively, consistent with the expected $\sqrt{2}$ pulse broadening per SHG stage.

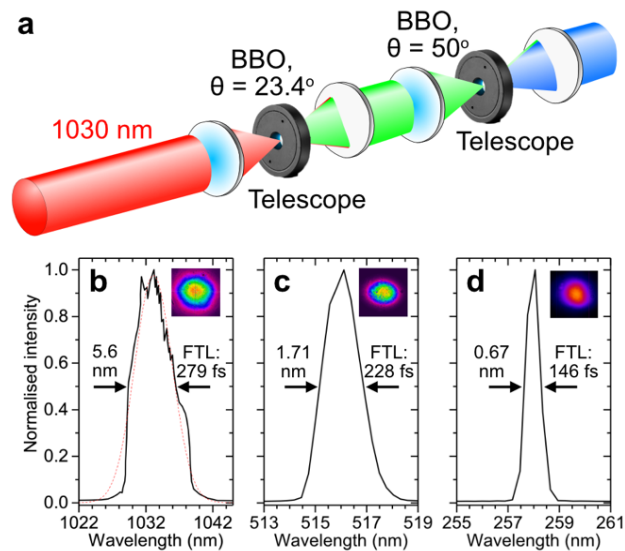

**FIG. 5.** Fourth harmonic generation (FHG) of ultraviolet (UV) radiation. (a) Schematic of cascaded FHG setup. (b-d) Spectra and beam profiles of 1030 nm, 515 nm, and 257.5 nm beams, with their FWHM bandwidth and Fourier-transform limited (FTL) pulse duration indicated. Gaussian fit to spectrum is shown as red dotted distributions.

Figure 6 summarizes FHG from 40 kHz to 600 kHz. The UV output increases at low infrared (IR) power and saturates near 55 mW (see Fig. 6a). The maximum UV pulse energy decreases approximately an order of magnitude per step in repetition rate (see Fig. 6b), yielding 10-100 nJ at 40 kHz, >10 nJ at 100 kHz, >2 nJ at 200 kHz, and >50 pJ at 600 kHz. This energy range matches UED photoemission requirements, where 1 pJ - 10 nJ is sufficient to generate electron bunches containing $1 – 10^6$ electrons.[8] Even at 200 kHz and 600 kHz, usable bunch charges (>$10^5$ and >5,000 electrons, respectively[8]) are achievable, demonstrating the feasibility of high-repetition rate UED operations up to 600 kHz, which is particularly relevant for UED applications with an electron beam containing more than 100 electrons/pulse. A maximum FHG conversion efficiency exceeding 10% is achieved at 40 kHz, decreasing to <3%, <1%, and <0.1% at 100 kHz, 200 kHz, and 600 kHz, respectively. The consistent efficiency maximum at ~55 mW of IR power across all repetition rates indicates that thermal effects are limiting phase matching in the BBO crystals. At higher powers, back-conversion between fundamental and harmonic radiation reduces the conversion efficiency, as expected in high-intensity SHG and FHG processes where strong harmonic buildup induces energy reconversion.[49]

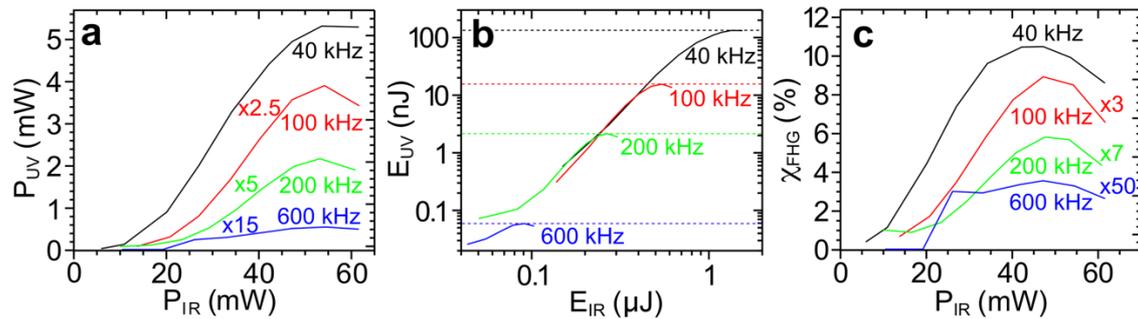

**FIG. 6.** Emitted ultraviolet (UV) radiation by fourth harmonic generation (FHG). (a) Measured UV power as a function of 1030 nm infrared (IR) power at different repetition rates. (b) Corresponding UV pulse energy as a function of IR pulse energy. (c) FHG conversion efficiency as a function of IR power.

This cascaded FHG concept offers a compact, low-complexity route to generating UV pulses directly from a Yb laser platform, particularly when operating at a fundamental wavelength of 1030 nm. The demonstrated lower-energy FHG stage supports stable nJ-pJ pulses suitable for photoemission in UED. Extending this FHG concept to µJ-level UV generation is possible with careful thermal management and optimization of beam size, focusing, group delay dispersion, and BBO crystal thickness at higher-energy fundamental pulses – a subject for future work.

## Conclusion

In summary, we have demonstrated scalable, high-repetition-rate generation of both THz and UV radiation driven by a Yb-based femtosecond laser system. Optical rectification in $LiNbO_3$ up to 100 kHz and cascaded FHG to 600 kHz confirm that efficient nonlinear conversion is achievable at repetition rates two-to-three orders of magnitude higher than in most UED systems which utilize more than 100 electrons/pulse as the probe electron beam. The THz fields provide the field strength for temporal compression and streaking, while the UV pulses deliver the necessary photon energies needed for photoemission and excitation. With appropriate thermal management and crystal optimization, extension to µJ-level UV generation is within reach. These results outline a clear pathway toward next-generation, high-throughput UED systems operating at tens-to-hundreds of kilohertz with sub-100 fs resolution and atto-to-pico-Coulomb-level bunch charges.

## Acknowledgements

We acknowledge financial support from the European Research Council for ERC Starting Grant "TERES" (Grant No. 101165245) and Lasers4EU (Grant No. 101131771, Project ID 37011). We are grateful to Peter Baum, Joel Kuttruff, Roman Peslin, Wolfgang Krueger, Michael Woerner, Thomas Elsaesser, Arnaud Rouzèe, and Marc J. J. Vrakking for support.

## AUTHOR DECLARATIONS

### Conflict of interest

The authors have no conflicts to disclose.

### Author contributions

Andrey Ryabov and Kasra Amini contributed to all aspects of this work.

## DATA AVAILABILITY

Data underlying the results presented in this paper are not publicly available at this time but may be obtained from the authors upon reasonable request.